%% file: jnl-2015-mtt-cim_so.tex
\documentclass{article}
\usepackage{tikz}
\usetikzlibrary{calc}
\usetikzlibrary{arrows}
 \usepackage{relsize}
 \usepackage[americanvoltage]{circuitikz}
\usepackage{geometry}
\usepackage{amsmath}
\usepackage{amssymb}
\usepackage{cite}
\usepackage{esint}
\usetikzlibrary{calc}
\usepackage[americanvoltage]{circuitikz}
\usepackage{paralist}

\input{myDefinitions}

\begin{document}

\title{Skin Effect Modeling in Conductors of Arbitrary Shape Through a Surface Admittance Operator and the Contour Integral Method}

\author{Utkarsh~R.~Patel
        and~Piero~Triverio  \thanks{This work was supported in part by the Natural Sciences and Engineering Research Council of
Canada (Discovery Grant program) and in part by the Canada Research Chairs program.}
\thanks{U.~R.~Patel and P.~Triverio are with the Edward S. Rogers Sr. Department of Electrical and Computer Engineering, University of Toronto, Toronto, M5S 3G4 Canada (email: utkarsh.patel@mail.utoronto.ca, piero.triverio@utoronto.ca).}\\[2em]
Submitted to IEEE Transactions on Microwave Theory and Techniques \\
on September 27, 2015
}

\date{}


\maketitle

\input{00abstract.tex}
\input{Introduction.tex}

\input{ProblemDefinition.tex}
\input{CIM.tex}

\input{SurfaceAdmittanceOperator.tex}
\input{NumericalResults.tex}

\input{Conclusion.tex}

\bibliographystyle{IEEEtran}
\bibliography{IEEEabrv,biblio}

\end{document}

%% file: myDefinitions.tex
\newcommand{\vect}[1]{\mathbf{#1}}

\newcommand{\matr}[1]{\mathbf{#1}}
\newcommand{\abs}[1]{\left \lvert #1 \right\rvert }
 
\newcommand{\pref}[1]{(\ref{#1})}
\newcommand{\junk}[1] {}

\def\XXint#1#2#3{{\setbox0=\hbox{$#1{#2#3}{\int}$}
\vcenter{\hbox{$#2#3$}}\kern-.5\wd0}}

\newcommand*\widebar[1]{%
  \hbox{%
    \vbox{%
      \hrule height 0.5pt 
      \kern0.3ex
      \hbox{%
        \kern-0.05em
        \ensuremath{#1}%
        \kern-0.05em
      }%
    }%
  }%
}


%% file: 00abstract.tex
\begin{abstract}
An accurate modeling of skin effect inside conductors is of capital importance to solve transmission line and scattering problems. This paper presents a surface-based formulation to model skin effect in conductors of arbitrary cross section, and compute the per-unit-length impedance of a multiconductor transmission line. The proposed formulation is based on the Dirichlet-Neumann operator that relates the longitudinal electric field to the tangential magnetic field on the boundary of a conductor. We demonstrate how the surface operator can be obtained through the contour integral method for conductors of arbitrary shape. The proposed algorithm is simple to implement, efficient, and can handle arbitrary cross-sections, which is a main advantage over the existing approach based on eigenfunctions, which is available only for canonical conductor's shapes. The versatility of the method is illustrated through a diverse set of examples, which includes transmission lines with trapezoidal, curved, and V-shaped conductors. Numerical results demonstrate the accuracy, versatility, and efficiency of the proposed technique.
\end{abstract}


%% file: Introduction.tex
\section{Introduction}

Transmission line modeling is a crucial part of the computer aided design of a variety of systems, including high-speed electronic boards~\cite{Young2001,Pau07}, integrated circuits~\cite{caignet2001challenge}, microwave systems~\cite{hong2004microstrip}, metamaterials~\cite{caloz2005electromagnetic}, and power grids~\cite{Ame13}. In order to create a transmission line model, one must first obtain the per-unit-length (p.u.l.) impedance and admittance of the line~\cite{Paul2011}. In several applications, such as the investigation of signal integrity issues in high-speed electronic systems, such parameters are required over a wide band, extending from DC up to tens of gigahertz~\cite{Young2001}. Across such band, the development of skin effect significantly changes the line behaviour~\cite{Pau07}, and must be accurately described. Skin effect modeling is also important in scattering problems~\cite{Tsang_2000}.

In order to obtain the p.u.l. impedance of a transmission line, one must describe the electromagnetic fields both inside the conductors (interior problem) and outside (exterior problem). For the exterior problem, integral equations~\cite{kamon1994fasthenry,Har61,Rue13} are commonly utilized. By requiring only a discretization of the conductors' surface, integral equations are typically more efficient that volumetric approaches such as the finite difference or finite element method (FEM)~\cite{jin2014finite,Cri89}, which require the discretization of a large area around the conductors.

For the interior problem, which is responsible for capturing skin effect, both volumetric and surface methods have been used. Volumetric approaches include the FEM~\cite{costache1987finite,Cri89}, conductor partitioning~\cite{Com73} and volumetric integral equations~\cite{Coperich_2001}. Unfortunately, as frequency increases, these methods become computationally inefficient, since a very fine mesh is needed to capture the pronounced skin effect.

Surface methods solve the interior problem using only a discretization of the boundary of the conductors. This is achieved by describing the electromagnetic behaviour of the conductor through a surface operator that relates the electric and magnetic fields on the boundary. This operator is responsible for modeling  skin  and proximity effects~\cite{Pau07}. The surface operator 
can be obtained analytically or numerically.
In~\cite{Tsuk1991,Al-Qedra_2009}, a surface operator is derived using the so-called surface-impedance boundary conditions~\cite{YuferevIda_2009}, under the approximation of small curvature and well-developed skin effect~\cite{Qian_2007}. While high-order boundary conditions have been proposed to increase accuracy~\cite{Rienzo2008}, this approach is mostly suitable for conductors with smooth boundaries.
Numerically, the surface operator can be obtained using finite differences~\cite{Cop00}, finite elements~\cite{Siripuram_2009}, the electric field integral equation~\cite{Menshov_2013,Tong_2014} or the magnetic field integral equation~\cite{Qian_2007}. All these approaches require, to calculate the surface operator, a volumetric discretization of the interior problem and the calculation of several kernel matrices whose size depends on mesh size. These features increase their computational cost and complexity.

An interesting technique to derive the surface operator analytically was introduced by Knockaert and De Zutter in~\cite{DeZ05}. Using the eigenfunctions of the Helmholtz equation, the surface admittance operator is obtained analytically, avoiding a discretization of the interior problem. This key idea leads to a simple and efficient formulation. However, since  eigenfunctions can realistically be obtained only for canonical geometries, this approach has been restricted so far to circular~\cite{DeZ05,TPWRD1}, tubular~\cite{TPWRD2}, rectangular~\cite{DeZ05}, and triangular conductors~\cite{Demeester_2010}.

In this paper, we show how the surface admittance operator for conductors of \emph{arbitrary} shape can be efficiently obtained from a simple contour integral, through the so-called contour integral method~\cite{Okoshi1985}. While the possibility to extract the surface admittance operator from the contour integral is already known~\cite{Knockaert2008}, it has been applied only to simple scattering problems~\cite{Knockaert2008}. All works that followed adopted the eigenfunctions approach~\cite{DeZ05, Zutter2007,Deemester_2008, Demeester_2010, Demeester_2010_2} which is feasible for canonical shapes only. While arbitrary shapes can be decomposed into triangles~\cite{Demeester_2010}, this significantly adds to the algorithm complexity. This paper provides a simple,  unified way to handle arbitrary cross-sections, and extract accurate and broadband p.u.l. parameters for a variety of transmission lines. Numerical tests show that the proposed method is robust and computationally efficient, even when compared against the analytical approach based on eigenfunctions~\cite{DeZ05}. In this paper, the proposed method is applied to calculate transmission line parameters, but it can also be used for scattering problems~\cite{Rogier2007}.

The paper is organized as follows. In Sec.~\ref{sec:ProblemDefinition}, we define the problem, and in Sec.~\ref{sec:TheoreticalFormulation} we discuss  how the surface admittance operator is related to the contour integral method from a theoretical standpoint. In Sec.~\ref{sec:NumericalFormulation}, we discuss the numerical implementation of the proposed approach, and in Sec.~\ref{sec:EFIE} how it can be used to compute the p.u.l. impedance of arbitrary transmission lines. Finally, in Sec.~\ref{sec:Results}, we demonstrate the accuracy, robustness, and computational efficiency of the proposed method through a comprehensive set of examples.

%% file: ProblemDefinition.tex
\section{Problem Definition}
\label{sec:ProblemDefinition}
We consider a system of $P$ conductors of arbitrary shape, having conductivity $\sigma$, permittivity $\varepsilon$, and permeability $\mu$.  For the sake of generality, the conductors are assumed to be inside a stratified medium, where each layer has permittivity $\varepsilon_l$ and permeability $\mu_l$. If a conductor extends into two layers, it is decomposed into two parts, as discussed in~\cite{Olyslager1993}. Our final goal is to calculate the partial p.u.l. resistance $\matr{R}(\omega)$ and inductance $\matr{L}(\omega)$ matrices in the Telegrapher's equation~\cite{Pau07}
\begin{equation}
\frac{\partial \vect{V}}{\partial z} = - \left[\matr{R}(\omega) + j\omega\matr{L}(\omega) \right] \vect{I}\,.
\label{eq:Telegrapher}
\end{equation}
In the equation above, $\vect{V} = \begin{bmatrix} V_1 & \hdots & V_P \end{bmatrix}^T$ collects the potential $V_p$ of each conductor, while $\vect{I} = \begin{bmatrix} I_1 & \hdots & I_P \end{bmatrix}^T$ contains the current $I_p$ flowing in each conductor. 
The impedance parameters will be calculated with the surface admittance approach of~\cite{DeZ05} using, however, a contour integral to obtain the surface admittance operator.

%% file: CIM.tex
\section{Surface Admittance Operator Through the Contour Integral Method}
\label{sec:TheoreticalFormulation}

\begin{figure}[t]
\centering
\input{./Tikzfigure/fig1.tex}
\input{./Tikzfigure/fig2.tex}
\caption{Left panel: sample geometry of a conductor with arbitrary cross-section inside a stratified medium.
Right panel: equivalent configuration obtained after replacing the conductor with the surrounding medium, and introducing an equivalent current density $J_s^{(p)}(\vec{r})$ on contour $\gamma^{(p)}$.}
\label{fig:1}
\end{figure}
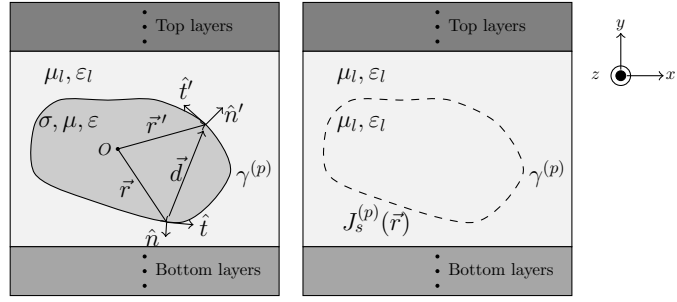

\subsection{Surface Admittance Formulation}

We discuss the surface admittance formulation used to solve the interior problem by considering the $p$-th conductor of a multiconductor transmission line. The conductor has arbitrary cross section, has a simply-connected contour $\gamma^{(p)}$, and is depicted in the left panel of Figure~\ref{fig:1}. The position vector of an arbitrary point on $\gamma^{(p)}$ is denoted by $\vec{r}$.
Finally, the unit vectors normal and tangential to contour $\gamma^{(p)}$ are denoted by $\hat{{n}}$ and $\hat{{t}}$, respectively, as shown in Fig.~\ref{fig:1}.

In order to derive the surface admittance operator, we initially follow~\cite{DeZ05}, and consider a boundary value problem.
Suppose that the $z$-directed electric field on contour $\gamma^{(p)}$ is given by ${E}_z^{(p)}(\vec{r})$.
Under the quasi-TM assumption~\cite{DeZ05}, the electric field ${\cal E}_z^{(p)}$ inside contour $\gamma^{(p)}$ can be obtained from the scalar Helmholtz equation~\cite{Bal89}
\begin{equation}
\nabla^2 {\cal E}_z^{(p)}+ k^2 {\cal E}_z^{(p)} = 0
\label{eq:Helmholtz1}
\end{equation}
subject to the boundary condition
\begin{equation}
{\cal E}_z^{(p)}(\vec{r}) = E_z^{(p)}(\vec{r}) \quad \quad \vec{r} \in \gamma^{(p)} \,.
\label{eq:Dirichlet1}
\end{equation}
In~\pref{eq:Helmholtz1}, $k = \sqrt{\omega \mu \left(\omega \varepsilon - j \sigma \right)}$ is the wavenumber inside the conductor.
Furthermore, the tangential magnetic field $H_t^{(p)} ({\vec{r}})$
along the contour $\gamma^{(p)}$ follows directly from Maxwell's equations under the quasi-TM assumption~\cite{Bal89} 
\begin{equation}
H_t^{(p)}({\vec{r}}) = \frac{1}{j \omega \mu} \left[ \frac{\partial {\cal E}_z^{(p)}({\vec{r}})}{\partial n} \right]_{\vec{r} \in \gamma^{(p)}}\\,
\label{eq:H1}
\end{equation}
where the derivative is taken along the direction normal to contour $\gamma^{(p)}$, as shown in the left panel of Fig.~\ref{fig:1}.

Next, we replace the conductor by the material of the surrounding layer~\cite{DeZ05}, as shown in the right panel of Fig.~\ref{fig:1}.
From here onwards, we will call this the equivalent configuration.
Due to this modification, the electric field inside the conductor changes to $\widetilde{\cal E}_z^{(p)} (\vec{r})$ which satisfies the Helmholtz equation
\begin{equation}
\nabla^2 \widetilde{\cal E}_z^{(p)} + k_{\text{out}}^2 \widetilde{\cal E}_z^{(p)} = 0
\label{eq:WaveEqn2}
\end{equation}
subject to the same Dirichlet boundary condition~\pref{eq:Dirichlet1}
\begin{equation}
\widetilde{\cal E}_z^{(p)} ({\vec{r}}) = E_z^{(p)} ({\vec{r}}) \quad \quad \vec{r} \in \gamma^{(p)}\,.
\label{eq:Dirichlet2}
\end{equation}
In~\pref{eq:WaveEqn2}, $k_{\text{out}} = \omega \sqrt{\mu_l \varepsilon_l}$ is the wavenumber inside the layer surrounding the conductor.
In this new configuration, the tangential magnetic field along $\gamma^{(p)}$ is
\begin{equation}
\widetilde{H}_t^{(p)}({\vec{r}}) = \frac{1}{j \omega \mu_l} \left[ \frac{\partial \widetilde{\cal E}_z^{(p)} ({\vec{r}})}{\partial n}\right]_{\vec{r} \in \gamma^{(p)}}\,.
\label{eq:H2}
\end{equation}

By replacing the conductor with the surrounding medium, we have modified the fields both inside and outside $\gamma^{(p)}$.
Hence, to restore the original fields outside $\gamma^{(p)}$, we invoke the equivalence principle~\cite{Bal05} and introduce an equivalent surface current  density $J_s^{(p)}(\vec{r})$ on $\gamma^{(p)}$~\cite{DeZ05}, as shown in the right panel of Fig.~\ref{fig:1}.
The equivalent current is directed along $z$, and is given by
\begin{equation}
J_s^{(p)}(\vec{r}) = {H}_t^{(p)} (\vec{r}) - \widetilde{H}_t^{(p)} (\vec{r})\,.
\label{eq:Js}
\end{equation}
It is important to note that an equivalent \emph{magnetic} current on $\gamma^{(p)}$ is not required, since the electric field on $\gamma^{(p)}$ in the original and equivalent configuration remain the same due to the Dirichlet conditions~\pref{eq:Dirichlet1} and \pref{eq:Dirichlet2}.
By substituting \pref{eq:H1} and \pref{eq:H2} into \pref{eq:Js}, we obtain~\cite{DeZ05}
\begin{equation}
J_s^{(p)}(\vec{r}) = \frac{1}{j \omega} \left[\frac{1}{\mu} \frac{\partial {\cal E}_z^{(p)}({\vec{r}})}{\partial n} - \frac{1}{\mu_l} \frac{\partial \widetilde{\cal E}_z^{(p)}({\vec{r}})}{\partial n} \right]_{\vec{r} \in \gamma^{(p)}}\,.
\label{eq:SAO}
\end{equation}
Equation~\eqref{eq:SAO} defines a surface admittance operator ${\cal Y}_s^{(p)}$ that relates the longitudinal electric field and equivalent current on $\gamma^{(p)}$ as
\begin{equation}
J_s^{(p)}(\vec{r}) = {\cal Y}_s^{(p)} E_z^{(p)}(\vec{r})\,.
\label{eq:SAO2}
\end{equation}
An explicit expression for ${\cal Y}_s^{(p)}$ can be written in terms of the eigenfunctions of the Helmhotz equations~\pref{eq:Helmholtz1} and~\pref{eq:WaveEqn2}, as shown in~\cite{DeZ05}. However, this approach is viable only for canonical conductor shapes, for which eigenfunctions are known analytically
\cite{DeZ05,Demeester_2010,TPWRD1,TPWRD2}. For arbitrary shapes, eigenfunctions can only be computed numerically. Since many eigenfunctions are needed to accurately model the operator, this approach can be very time consuming, and is typically avoided. In the next section, we show that the contour integral method~\cite{Okoshi1985} provides an efficient and robust way to compute such operator numerically for arbitrary shapes.

\subsection{Contour Integral Method}

Equating~\eqref{eq:SAO} and~\eqref{eq:SAO2}, we see that, in order to derive an explicit expression for the surface admittance operator, we need a relation between the electric field $E_z^{(p)}$ on the boundary and its normal derivative $\partial {\cal E}_z^{(p)} / \partial n$. With this goal in mind, we  reconsider the way we solve~\pref{eq:Helmholtz1}. The contour integral method~\cite{Okoshi1985} gives the solution of the Helmholtz equation in terms of the electric field and its normal derivative on $\gamma^{(p)}$
\begin{align}
{\cal E}_z^{(p)}(\vec{r}) = \frac{j}{2} \ointctrclockwise_{\gamma^{(p)}} \Bigg [&\frac{\partial G({\vec{r}},\vec{r}\,')}{\partial n'} {\cal E}_z^{(p)}({\vec{r}\,'}) \nonumber \\
&- G(\vec{r}, \vec{r}\,')\frac{\partial {\cal E}_z^{(p)} (\vec{r}\,')}{\partial n'} \Bigg ] dr'
\label{eq:CIM}
\end{align}
where $\vec{r}$ and $\vec{r}\,'$ are both on $\gamma^{(p)}$, 
and $n'$ is the unit vector normal to the contour at the point $\vec{r}\,'$.
The Green's function is
\begin{equation}
G(\vec{r}, \vec{r}\,')  = C_0 J_0 (k d) - jY_0 (k d)
\label{eq:CIM_Green}
\end{equation}
where $J_0(.)$ and $Y_0(.)$ are the zero-th order Bessel and Neumann functions~\cite{Abr64}, respectively. As shown in the left panel of Fig.~\ref{fig:1}, the distance between points $\vec{r}$ and $\vec{r}\,'$ is denoted as
\begin{equation}
\vec{d} = \vec{r}\,' - \vec{{r}}\,,
\label{eq:R1}
\end{equation}
and $d = {\abs{\vec{d}\,}}$.
Equation~\pref{eq:CIM} stems from the planar Green's theorem, and shows that the electric field along the contour $\gamma^{(p)}$ can be interpreted as the superposition of cylindrical waves that originate from points along $\gamma^{(p)}$.
Constant $C_0$ can be any complex number. If $C_0 = 1$ the Green's function is the Hankel function of the second kind, which represents outgoing cylindrical waves. In Sec.~\ref{sec:NumericalIssues}, we will discuss how to choose $C_0$ at low and high frequency to achieve high numerical robustness. 
The contour integral equation~\eqref{eq:CIM} relates the electric field on the boundary to its normal derivative. It can thus be used, after numerical discretization, to derive an explicit expression for the surface admittance operator~\eqref{eq:SAO2}.

%% file: Tikzfigure/fig1.tex
\usetikzlibrary{arrows, decorations.markings}

\begin{tikzpicture}[scale=0.65, every node/.style={scale=0.65}]
\draw [fill = black!50] (-1,1.5) rectangle (4.5,2.5);
\draw [fill = black!5] (-1,-2.5) rectangle (4.5,1.5);
\draw [fill = black!35] (-1,-2.5) rectangle (4.5,-3.5);
\draw [black, fill = black!20] plot [smooth cycle] coordinates {(-0.5,-1) (-0.5,0) (0,0.5) (1,0.5) (2,0.5) (3,0) (3.5,-1) (2.5,-2) (0.5,-1.5)};
\node [right] at (3.5,-1) {\Large $\gamma^{(p)}$};

\draw[fill=black]  (1.75,1.7) ellipse (0.03 and 0.03);
\draw[fill=black]  (1.75,2.0) ellipse (0.03 and 0.03);
\draw[fill=black]  (1.75,2.3) ellipse (0.03 and 0.03);
\draw[fill=black]  (1.75,-2.7) ellipse (0.03 and 0.03);
\draw[fill=black]  (1.75,-3) ellipse (0.03 and 0.03);
\draw[fill=black]  (1.75,-3.3) ellipse (0.03 and 0.03);

\node at (1.85, 2.0) [right] {Top layers};
\node at (1.85, -3.0) [right] {Bottom layers};



\node (p1) at (3,0) {};
\node (p2) at (3.35,0.35) {};
\node (p3) at (2.55,0.4) {};

\draw [->] (p1.center) -- (p3.center) {};
\draw [->] (p1.center) -- (p2.center) {};
\node (p4) at (3.25, 0.25) [above,right] {\Large $\hat{n}'$};
\node (p5) at (2.6,0.4) [above] {\Large $\hat{t}'$};

\draw [fill = black]  (1.2,-0.5) ellipse (0.03 and 0.03);
\node at (1.2,-0.5) [left] { $O$};
\node (p6) at (1.2, -0.5) {};

\node (p7) at (3,0) {};
\node (p8) at (2.2,-2) {};
\draw [->] (p6.center) -- (p7.center) {};
\draw [->] (p6.center) -- (p8.center) {};
\draw[->] (p8.center) -- (2.18, -2.3){};
\draw[->] (p8.center) -- (2.75, -2.05){};
\node at (2.18,-2.3) [left] {\Large $\hat{n}$};
\node at (2.75,-2.05) [right] {\Large $\hat{t}$};

\node at (2,-0.3) [above] {\Large $\vec{r}\,'$};
\node [below] at (1.6,-1.3)[left] {\Large $\vec{{{r}}}$};
\draw [->] (p8) -- (p7) {};
\node at (2.65,-0.9) [left] {\Large $\vec{d}$};

\begin{scope}[shift={(-0.5,-1)}]
\node (p9) at (4,2) {};
\node (p10) at (5,2) {};
\node (p11) at (4,3) {};
\end{scope}

\node at (0.2,0) {\Large $\sigma, \mu, \varepsilon$};
\node at (0.2,1) {\Large $\mu_l, \varepsilon_l$};

\end{tikzpicture}

%% file: Tikzfigure/fig2.tex
\begin{tikzpicture}[scale=0.65, every node/.style={scale=0.65}]

\draw [fill = black!50] (-1,1.5) rectangle (4.5,2.5);
\draw [fill = black!5] (-1,-2.5) rectangle (4.5,1.5);
\draw [fill = black!35] (-1,-2.5) rectangle (4.5,-3.5);
\draw[fill=black]  (1.75,1.7) ellipse (0.03 and 0.03);
\draw[fill=black]  (1.75,2.0) ellipse (0.03 and 0.03);
\draw[fill=black]  (1.75,2.3) ellipse (0.03 and 0.03);
\draw[fill=black]  (1.75,-2.7) ellipse (0.03 and 0.03);
\draw[fill=black]  (1.75,-3) ellipse (0.03 and 0.03);
\draw[fill=black]  (1.75,-3.3) ellipse (0.03 and 0.03);
\node at (1.85, 2.0) [right] {Top layers};
\node at (1.85, -3.0) [right] {Bottom layers};

\draw [black, dashed] plot [smooth cycle] coordinates {(-0.5,-1) (-0.5,0) (0,0.5) (1,0.5) (2,0.5) (3,0) (3.5,-1) (2.5,-2) (0.5,-1.5)};
\node [right] at (3.5,-1) {\Large $\gamma^{(p)}$};

\node (p1) at (3,0) {};
\node (p2) at (3.35,0.35) {};
\node (p3) at (2.55,0.4) {};

\node (p6) at (1.2, -0.5) {};

\begin{scope}[shift={(1.5,-1)}]

\draw  (4,2) ellipse (0.2 and 0.2);
\draw [fill = black] (4,2) ellipse (0.1 and 0.1);
\node (p9) at (4,2) {};
\node (p10) at (5,2) {};
\node (p11) at (4,3) {};
\draw [->] (p9) -- (p10) {};
\draw[->] (p9) -- (p11) {};
\node at (3.7,2) [left] {$z$};
\node at (4,2.8) [above] {$y$};
\node at (4.8,2) [right] {$x$};
\end{scope}

\node at (0.2,0) {\Large $\mu_l, \varepsilon_l$};
\node at (0.2,1) {\Large $\mu_l, \varepsilon_l$};

\node at (0.5,-1.5) [below] {\Large $J_s^{(p)}({\vec{r}})$};
\end{tikzpicture}

%% file: SurfaceAdmittanceOperator.tex
\section{Numerical Formulation}
\label{sec:NumericalFormulation}

\subsection{Discretization of Electric Fields, Magnetic Fields, and Equivalent Current}

We discretize the contour integral equation with the method of moments~\cite{Har61}, using point matching to test the resulting equation. We divide contour $\gamma^{(p)}$ into $N_p$ segments $\gamma_n^{(p)}$, and expand the longitudinal electric field in terms of pulse basis functions
\begin{equation}
E_z^{(p)}(\vec{r}) = \sum_{n=1}^{N_p} e_n^{(p)} \Pi_n^{(p)}(\vec{r})\,,
\label{eq:E_expansion}
\end{equation}
where $\Pi_n^{(p)}(\vec{r})$ is the $n$-th pulse basis function which is one if $\vec{r}$ belongs to the $n$-th segment, and zero otherwise.
We also define $\vec{r}_n$ to be the position vector of the midpoint of the $n$-th partition.
Similarly, we discretize the tangential magnetic fields ${H}_t^{(p)}(\vec{r})$ and $\widetilde{H}_t^{(p)}(\vec{r})$ that appear in \pref{eq:Js} as
\begin{equation}
H_t^{(p)}(\vec{r}) = \sum_{n=1}^{N_p} h_n^{(p)} \Pi_n^{(p)}(\vec{r})\,,
\label{eq:H_expansion}
\end{equation}
and 
\begin{equation}
\widetilde{H}_t^{(p)}(\vec{r}) = \sum_{n=1}^{N_p} \tilde{h}_n^{(p)} \Pi_n^{(p)}(\vec{r})\,.
\end{equation}
For simplicity of notation, we cast all expansion coefficients into column vectors
\begin{align}
\vect{E}^{(p)} & = \begin{bmatrix}e_1^{(p)} & e_2^{(p)} & \hdots & e_{N_p}^{(p)} \end{bmatrix}^T\,, \\
{\vect{H}}^{(p)} &= \begin{bmatrix} {h}_1^{(p)} & {h}_2^{(p)} & \hdots & {h}_{N_p}^{(p)} \end{bmatrix}^T \,, \\
\widetilde{\vect{H}}^{(p)} &= \begin{bmatrix} \tilde{h}_1^{(p)} & \tilde{h}_2^{(p)} & \hdots & \tilde{h}_{N_p}^{(p)} 
\end{bmatrix}^T\,.
\end{align}
Similarly, we discretize equivalent current $J_s^{(p)}(\vec{r})$ along the contour $\gamma^{(p)}$ using pulse basis functions as
\begin{equation}
J_s(\vec{r}) = \sum_{n=1}^{N_p} j_n^{(p)} \Pi_n^{(p)}(\vec{r})\,,
\label{eq:J_expansion}
\end{equation}
with coefficients $j_n^{(p)}$ stored into vector
\begin{equation}
\vect{J}^{(p)} = \begin{bmatrix} j_1^{(p)} & j_2^{(p)} & \hdots & j_{N_p}^{(p)} \end{bmatrix}^T\,.
\end{equation}
From~\pref{eq:Js}, we have the following relation between the coefficients of equivalent current and magnetic fields
\begin{equation}
\vect{J}^{(p)} = \vect{H}^{(p)} - \widetilde{\vect{H}}^{(p)}\,.
\label{eq:Js2}
\end{equation}
Next, we relate $\vect{H}^{(p)}$ and $\widetilde{\vect{H}}^{(p)}$ to the electric field coefficients $\vect{E}^{(p)}$ via the contour integral equation \pref{eq:CIM} to obtain the surface admittance operator.

\subsection{Magnetic Field in the Original Configuration}
\label{sec:Hfield_original}

After combining~\eqref{eq:H1} and~\eqref{eq:CIM}, we substitute the electric and magnetic field expansions~\pref{eq:E_expansion} and~\pref{eq:H_expansion} into the contour equation to obtain
\begin{align}
\sum_{m=1}^{N_p} e_m^{(p)} \Pi_m^{(p)}(\vec{r}) = \frac{j}{2} \ointctrclockwise_{\gamma^{(p)}} \Bigg [ \frac{\partial G(\vec{r},\vec{r}\,') }{\partial n'} \sum_{n=1}^{N_p} e_n^{(p)} \Pi_n^{(p)}(\vec{r}\,')  \nonumber \\
- j \omega \mu G(\vec{r}, \vec{r}\,') \sum_{n=1}^{N_p} h_n^{(p)} \Pi_n^{(p)}(\vec{r}\,') \Bigg ] dr' 
\label{eq:CIM2}
\end{align}
Using point-matching~\cite{Bal05}, we test the equation above at the midpoints $\vec{r}_m$ of all $N_p$ segments, obtaining
\begin{align}
e_m^{(p)} = \frac{j}{2}  \sum_{n=1}^{N_p} e_n^{(p)} \int_{\gamma_n^{(p)}} \frac{\partial G(\vec{r}_m,\vec{r}\,') }{\partial n'}   dr'  \nonumber \\
+ \frac{\omega \mu}{2}  \sum_{n=1}^{N_p} h_n^{(p)}  \int_{\gamma_n^{(p)}} G(\vec{r}_m, \vec{r}\,') dr' 
\label{eq:CIM3}
\end{align}
for $m = 1, \hdots, N_p$. Note that in~\pref{eq:CIM3}, the integration is performed only over the $n$-th segment $\gamma_n^{(p)}$.
All relations~\pref{eq:CIM3} can be compactly written in matrix form as
\begin{equation}
\matr{U} \vect{E}^{(p)} = \matr{P} \vect{H}^{(p)}\,,
\label{eq:EH_relation}
\end{equation}
where $\matr{U}$ and $\matr{P}$ are square matrices with dimension $N_p \times N_p$. Element $(m,n)$ of matrix $\matr{U}$, if $m \neq n$, is given by
\begin{align}
\left[ \matr{U} \right]_{m,n} = \frac{jk}{2} \int_{\gamma_n^{(p)}}  \frac{\vec{d}_m\cdot \hat{{n}}'}{d_m} \left[C_0 J_1(kd_m) - jY_1(kd_m) \right] dr'\,,
\label{eq:U_entry1}
\end{align}
where $\vec{d}_m = \vec{r}\,' - \vec{r}_m$, $d_m = {\abs{\vec{d}_m}}$, and we used the fact that the normal derivative of~\eqref{eq:CIM_Green} can be written as
\begin{equation}
\frac{\partial G({\vec{r}}, \vec{r}\,')}{\partial n'} = - k \frac{\vec{d}\cdot \hat{n}'}{d} \left[ C_0 J_1(kd) - j Y_1(kd)\right] \,.
\label{eq:CIM_Green_der}
\end{equation}
The diagonal entries of $\matr{U}$ are given by
\begin{align}
	\left[\matr{U}\right]_{m,m} = 1\,.
\label{eq:U_entry2}
\end{align} 
The $(m,n)$-th entry of $\matr{P}$ in \pref{eq:EH_relation} is given by
\begin{align}
\left[ \matr{P} \right]_{m,n} = \frac{\omega \mu }{2} \int_{\gamma_n^{(p)}} \left[ C_0 J_0(kd_m) - j Y_0(kd_m) \right] dr'\,.
\label{eq:P_entry1}
\end{align} 
In all numerical examples of Sec.~\ref{sec:Results}, the integrals in~\eqref{eq:U_entry1} and~\eqref{eq:P_entry1} were evaluated using a 5-point Gaussian quadrature routine.
Since the Neumann function approaches infinity for small arguments~\cite{Abr64}, the diagonal entries of $\matr{P}$ must be evaluated analytically using the small-argument approximation of the Bessel functions, as done in~\cite{Okoshi1985}. 

From~\pref{eq:EH_relation}, we can express the magnetic field $\vect{H}^{(p)}$ in terms of the electric field on the same conductor as
\begin{equation}
\vect{H}^{(p)} = \matr{P}^{-1} \matr{U} \vect{E}^{(p)}\,.
\label{eq:H1_discretized}
\end{equation}

\subsection{Magnetic Field in the Equivalent Configuration}

Next, we find the magnetic field in the equivalent configuration by 
repeating all the steps of Sec.~\ref{sec:Hfield_original}, but with the material parameters of the conductor replaced by the parameters of the surrounding medium. These steps lead to
\begin{equation}
\widetilde{\vect{H}}^{(p)} = \matr{P}_{\text{out}}^{-1} \matr{U}_{\text{out}} \vect{E}^{(p)}\,,
\label{eq:H2_discretized}
\end{equation}
which is analogous to~\pref{eq:H1_discretized}.
In the equation above, entries of $\matr{P}_{\text{out}}$ and $\matr{U}_{\text{out}}$ are calculated with formulas  \pref{eq:P_entry1}, \pref{eq:U_entry1}, and \pref{eq:U_entry2}, with wavenumber $k$ replaced by $k_{\text{out}}$ and permeability $\mu$ replaced by $\mu_{l}$.

\subsection{Surface Admittance Operator}

Finally, by substituting \pref{eq:H1_discretized} and \pref{eq:H2_discretized} into \pref{eq:Js2}, we obtain
\begin{equation}
\vect{J}^{(p)} = \matr{Y}^{(p)} \vect{E}^{(p)}
\label{eq:SAO3}
\end{equation}
where
\begin{equation}
\matr{Y}^{(p)} = \matr{P}^{-1} \matr{U} - \matr{P}_{\text{out}}^{-1} \matr{U}_{\text{out}}
\end{equation}
is the discretized surface admittance operator of the $p$-th conductor.
Therefore, we see that with the proposed contour integral approach, it is sufficient to evaluate matrices $\matr{P}$, $\matr{P}_{\text{out}}$, $\matr{U}$, and $\matr{U}_{\text{out}}$ to easily obtain the surface admittance operator for a conductor of arbitrary shape. 

In order to compact the notation, we collect the coefficients of the electric field and equivalent current of all conductors into two column vectors $\vect{E}$ and $\vect{J}$. The surface admittance relations~\eqref{eq:SAO3} can be thus compactly written as
\begin{equation}
\vect{J} = \matr{Y}_s \vect{E}\,,
\label{eq:SAO4}
\end{equation}
where
\begin{equation}
\matr{Y}_s = \begin{bmatrix}
\matr{Y}^{(1)} &  & & \\
& \matr{Y}^{(2)} & &  \\
& & \ddots &  \\
& & & \matr{Y}^{(P)}
\end{bmatrix}\,.
\end{equation}

\subsection{Choice of $C_0$}
\label{sec:NumericalIssues}

In this section, we discuss how to set the constant $C_0$  at low and high frequency in order to achieve a well-conditioned algorithm.

\subsubsection{Low Frequency}

At low frequency, where skin effect has not yet developed, computing~\pref{eq:P_entry1} requires the evaluation of the Green's function~\pref{eq:CIM_Green} for very small arguments.  For small arguments, the Neumann function $Y_0(.)$ dominates  the Bessel function $J_0(.)$~\cite{Abr64}. Hence, to maintain a good numerical contrast between the Bessel and Neumann functions, we must set $C_0$ to a high value, as discussed in~\cite{Ayasli1980}. The value of $10^6$ provided accurate results for all numerical tests we performed, including the examples of Sec.~\ref{sec:Results}.

\subsubsection{High Frequency}

At high frequency, we set $C_0 = 1$. This makes the Green's function~\eqref{eq:CIM_Green} become the Hankel function of the second kind, which is well-behaved at high frequency.

\subsubsection{Switching Condition}

As previously discussed, two different values of $C_0$ are appropriate at low and high frequency. 
Numerical tests demonstrated that there is a large range of intermediate frequencies where both values of $C_0$ provide accurate results. Therefore, the choice of the frequency where one should switch from one $C_0$ value to the other is not critical. In our implementation, we use the low-frequency value of $C_0$ on the $p$-th conductor when
\begin{equation}
\frac{\Delta_p}{\delta_p} \le t \,,
\label{eq:switch}
\end{equation}
where $\Delta_p$ and $\delta_p$ are, respectively, the minimum transversal dimension and the skin depth in the $p$-th conductor. In all numerical test we performed, any $t$ value between 0.2 and 0.5 gave good results, and  $t=0.5$ has been used in all numerical examples presented in Sec.~\ref{sec:Results}. When frequency increases and inequality~\eqref{eq:switch} no longer holds, the high-frequency value of $C_0$ is used.

\section{Exterior Problem and Impedance Computation}
\label{sec:EFIE}

In this section, we use the electric field integral equation to express the relation between the electric field on the conductors' boundary, and the equivalent currents dictated by the region outside the conductors~\cite{DeZ05}. Combined with the surface admittance operator~\eqref{eq:SAO4}, this will lead to the p.u.l. impedance of the transmission line. On the contour of the $p$-th conductor, the electric field integral equation~\cite{Bal89} reads
\begin{equation}
{E}_z^{(p)}(\vec{r}) = j\omega \mu_l \sum_{q=1}^{P} \ointctrclockwise_{\gamma^{(q)}} J_s^{(q)}(\vec{r}\,') G_0(\vec{r}, \vec{r}\,') ds'  - \frac{\partial V_p}{\partial z}
\label{eq:efie}
\end{equation}
where $\vec{r} \in \gamma^{(p)}$ and the integration is performed over the contour $\gamma^{(q)}$ of each conductor ($q=1,\dots,P$).
The integral kernel $G_0(\vec{r}, \vec{r}\,')$ is the Green's function of the surrounding medium. Since, through the surface admittance formulation, all conductors have been replaced with the surrounding medium, $G_0(\vec{r},\vec{r}\,')$ is simply the Green's function of a stratified medium~\cite{Chew_1995}.
Following~\cite{DeZ05,Coperich_2001}, we substitute~\pref{eq:Telegrapher}, \pref{eq:E_expansion} and~\pref{eq:J_expansion} into~\pref{eq:efie}, obtaining
\begin{align}
\sum_{m=1}^{N_p} e_m^{(p)} \Pi_m^{(p)}(\vec{r}) = &j \omega \mu_l \sum_{q=1}^{P} \ointctrclockwise_{\gamma^{(q)}} \sum_{n = 1}^{N_q} j_n^{(q)} \Pi_n^{(q)}(\vec{r}\,') G_0(\vec{r}, \vec{r}\,') ds_q' \nonumber\\
&+ \sum_{q=1}^{P}\left[  \matr{R}_{pq}(\omega) + j\omega \matr{L}_{pq}(\omega) \right] I_q
\label{eq:efie1}
\end{align}
for $p = 1, \hdots, P$. Using point-matching~\cite{Bal05}, this integral equation can be converted into the system of algebraic equations
\begin{equation}
\vect{E} = j\omega\mu_l \matr{G}_0 \vect{J} + \matr{Q} \left[ \matr{R}(\omega) + j \omega \matr{L}(\omega)\right] \vect{I}
\label{eq:efie2}
\end{equation} 
where $\matr{G}_0$ is the Green's matrix.
In~\pref{eq:efie2}, $\matr{Q}$ is a block diagonal matrix
\begin{equation}
\matr{Q} = \begin{bmatrix}
\vect{1}_{1} & & & \\
& \vect{1}_2 & & \\
& & \ddots & \\
& & & \vect{1}_P
\end{bmatrix}
\end{equation}
where $\vect{1}_p$ is a vector of size $N_p \times 1$ whose entries are all ones.
Furthermore, as shown in~\cite{DeZ05}, the total current inside each conductor is equal to the contour integral of the equivalent current \pref{eq:J_expansion}, and hence
\begin{equation}
\vect{I} = {\matr{Q}}^T \matr{W} \vect{J}\,
\label{eq:IQJ}
\end{equation}
where ${\matr{W}}$ is a diagonal matrix in the form
\begin{equation}
{\matr{W}} = \begin{bmatrix} 
\vect{w}^{(1)} & & & \\
& \vect{w}^{(2)} & & \\
& & \ddots & \\
& & & \vect{w}^{(P)}
\end{bmatrix}
\end{equation}
where $\vect{w}^{(p)}$ is a diagonal matrix of size of $N_p  \times N_p$, where entry $(n,n)$ is the width of the $n$-th segment of $\gamma^{(p)}$.

As shown in~\cite{TPWRD1}, we can manipulate~\pref{eq:efie2} using~\pref{eq:SAO4} and~\pref{eq:IQJ} to obtain the partial p.u.l. resistance
\begin{equation}
\matr{R}(\omega) = \left[ {\matr{Q}}^T \matr{W} \left(\matr{1} - j\omega \mu_l \matr{Y}_s \matr{G}_0 \right)^{-1} \matr{Q} \right]^{-1}
\end{equation}
and partial p.u.l. inductance
\begin{equation}
\matr{L}(\omega) = \omega^{-1} \left[ {\matr{Q}}^T \matr{W} \left(\matr{1} - j\omega \mu_l \matr{Y}_s \matr{G}_0 \right)^{-1} \matr{Q} \right]^{-1}\,.
\end{equation}
The p.u.l. impedance can be obtained from the partial p.u.l. impedance by taking one of the conductors as reference.

%% file: NumericalResults.tex
\section{Numerical Results}
\label{sec:Results}

In this section, we demonstrate the efficiency and robustness of the proposed method, and compare it against the state of the art. We illustrate the versatility of the technique by considering a comprehensive set of transmission lines made by circular, rectangular, trapezoidal, V-shaped, and conformal conductors. The method is validated against the eigenfunctions approach~\cite{DeZ05, TPWRD1}, as well as a completely independent technique (FEM)~\cite{COMSOL}.
All computations were performed on a computer with 16~GB of memory and a 3.4~GHz processor. All techniques based on a surface admittance operator were implemented in MATLAB.

\subsection{Two-conductor lines}
\label{sec:example_twoconductor}
\input{./TestCases/TwoRoundConductors/Twoconductor_geometry}
\subsubsection{Round conductors}
\label{sec:example_tworoundconductor}
\begin{figure}[t]
\centering
\includegraphics{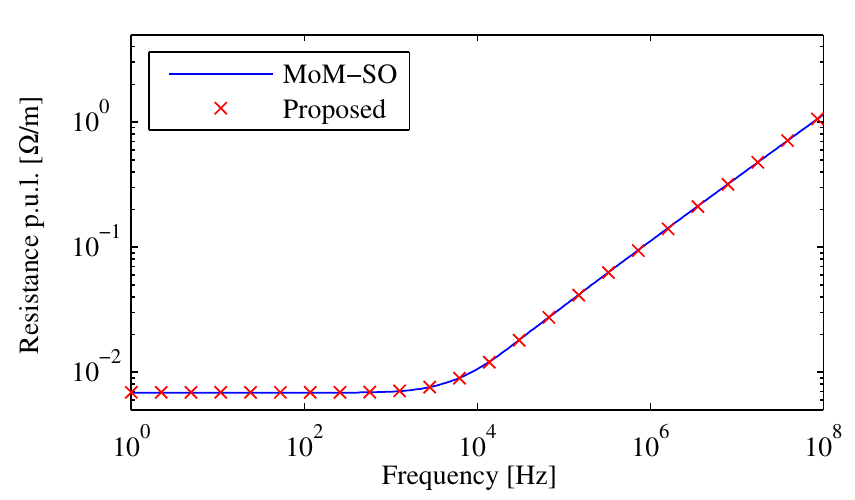}
\includegraphics{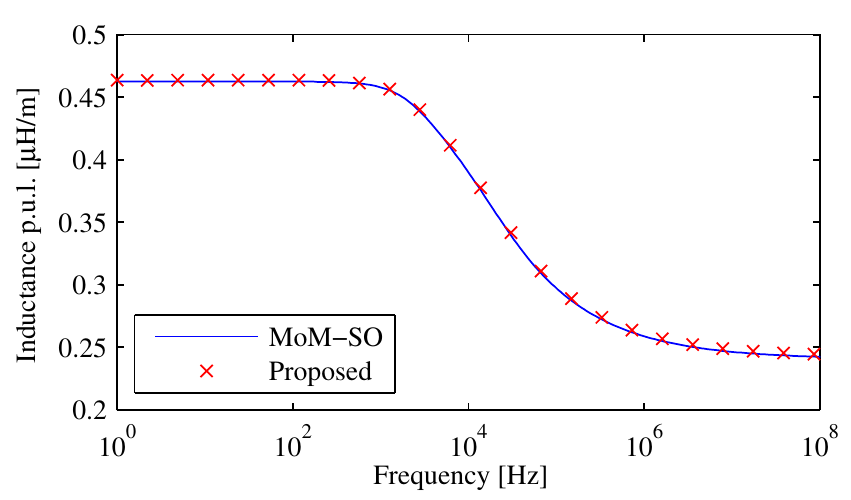}
\caption{P.u.l. resistance and inductance of the transmission line with two round conductors considered in Sec.~\ref{sec:example_tworoundconductor}, obtained with the proposed method and MoM-SO~\cite{TPWRD1}.}
\label{fig:Z_two_round_conductors}
\end{figure}

We first consider a transmission line made by two round conductors with radii $a_1 = 1~{\rm mm}$, $a_2 = 2~{\rm mm}$, and spacing $d = 3.5~{\rm mm}$. The line cross section is shown in the left panel of Fig.~\ref{fig:Twoconductor_geometry}. 
The conductivity of both conductors is $\sigma = 5.8 \cdot 10^7 ~{\rm S/m}$.
We calculated the p.u.l. impedance using the proposed technique and the MoM-SO algorithm~\cite{TPWRD1}, which uses the eigenfunctions method to derive the surface admittance operator. 
Figure~\ref{fig:Z_two_round_conductors} shows the p.u.l. resistance and inductance obtained with the proposed approach and with MoM-SO.
The two methods are in excellent agreement.
For the proposed approach, the boundaries enclosing the small and large conductors were discretized with $N_1 = 28$ and $N_2 = 60$ pulse basis functions, respectively.
As shown in Table~\ref{tab:Timing_twoconductor}, for each frequency point the proposed approach took only 0.04~s. In this case, the eigenfunctions approach is faster since it can capture, with few Fourier basis functions, the field distribution inside the circular conductors. The advantage of the proposed approach is generality: it can be applied to arbitrary shapes, while MoM-SO is limited to round conductors, either solid~\cite{DeZ05,TPWRD1} or hollow~\cite{TPWRD2}.

\begin{table}[!t]
\centering
\caption{Example of Sec.~\ref{sec:example_twoconductor}: CPU time required to compute the p.u.l. impedance at one frequency}
\begin{tabular}{|c|c|c|c|}
\hline
{\bf Test Case} & {\bf Proposed}  & {\bf MoM-SO} \cite{TPWRD1}  &  {\bf De Zutter} \cite{DeZ05}\\  \hline \hline
Round conductors&  0.04~s & 0.0006~s & N/A \\ \hline
Rectangular conductors& 0.16~s & N/A & 0.12~s \\ \hline
\end{tabular} \\
\label{tab:Timing_twoconductor} 
\end{table}

\begin{figure}[t]
\centering
\includegraphics{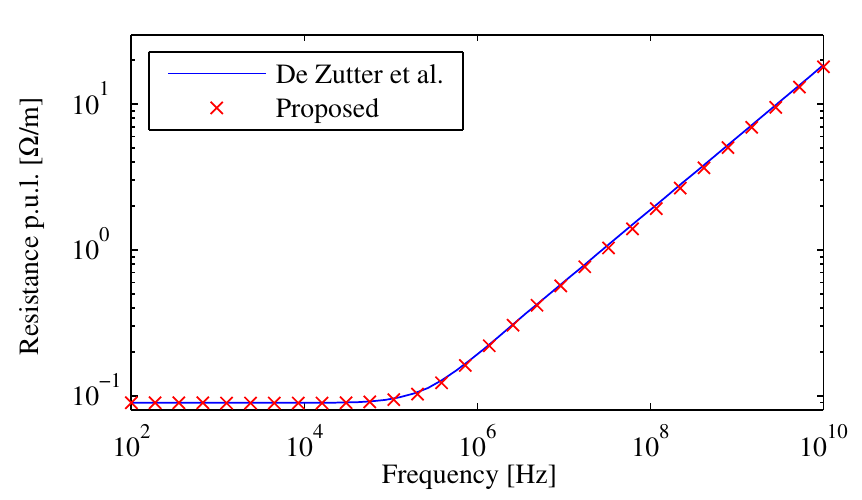}
\includegraphics{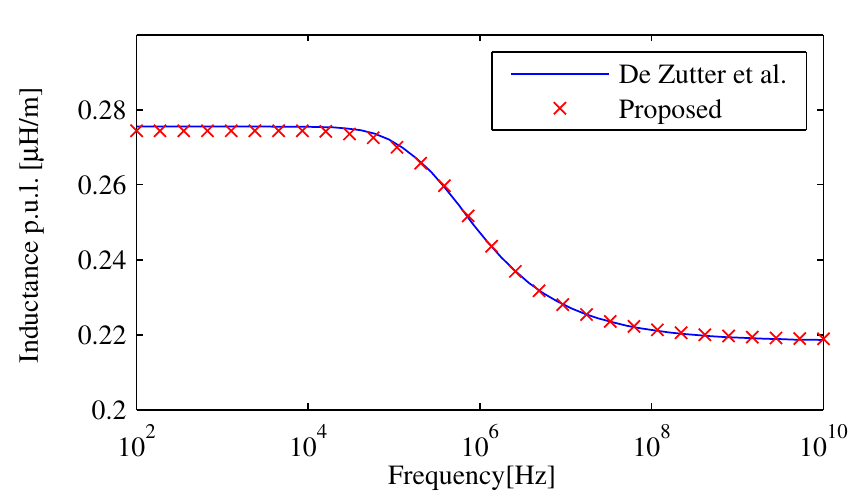}
\caption{P.u.l. resistance and inductance of the transmission line with two rectangular conductors considered in Sec.~\ref{sec:example_tworectangularconductor}. Results obtained with~\cite{DeZ05} are labeled as ``De Zutter et al''.}
\label{fig:Z_two_rectangular_conductors}
\end{figure}

\subsubsection{Rectangular conductors}
\label{sec:example_tworectangularconductor}
We now consider the case of two rectangular conductors presented in~\cite{DeZ05}.
Each conductor has conductivity $\sigma = 5.6\cdot 10^7~{\rm S/m}$ and dimension  $2~{\rm mm} \times 0.2~{\rm mm}$, as shown in the right panel of Fig.~\ref{fig:Twoconductor_geometry}. 
Figure~\ref{fig:Z_two_rectangular_conductors} shows the p.u.l. resistance and inductance for this transmission line computed with the proposed method and with~\cite{DeZ05}. The latter method computes the surface admittance operator analytically using sinusoidal functions, which are the eigenfunctions of  the Helmholtz equation on a rectangular domain. In order to interface the surface admittance operator to the electric field integral equation~\eqref{eq:efie1}, the sinusoidal functions have to be mapped onto pulse basis functions~\cite{DeZ05}. From Fig.~\ref{fig:Z_two_rectangular_conductors}, we see that the resistance and inductance values obtained with both approaches match very well.
For the proposed approach, each conductor was discretized with $N_1=N_2= 106$ pulse basis functions.
For the method of~\cite{DeZ05}, the number of sinusoidal harmonics were chosen to be $M = 400$ (see~\cite{DeZ05} for definition of $M$), and each conductor was discretized with $N_1 = N_2 = 280$ pulse basis functions.
The approach of~\cite{DeZ05} required more basis functions than the proposed approach to get accurate results at very high frequency.
The computational time taken by both techniques is given in Table~\ref{tab:Timing_twoconductor}. We see that the computational cost of the proposed approach is not far from the cost of~\cite{DeZ05}, although the proposed method is more general, as it can handle arbitrary shapes.

\subsection{Valley Microstrip Line}
\label{sec:valley}
\input{./TestCases/Valley/valley_geometry.tex}

\begin{table}[!t]
\centering
\caption{Examples of Sec.~\ref{sec:valley}, \ref{sec:Trapezoidal} and~\ref{Sec:Curved_Interconnect}: CPU time required to compute the p.u.l. impedance at one frequency with the proposed technique and the FEM~\cite{COMSOL}.}
\begin{tabular}{|c|c|c|c|}
\hline
{\bf Section} & {\bf Test Case} & {\bf Proposed}  & {\bf FEM}   \\  \hline \hline
Sec.~\ref{sec:valley} & Valley microstrip &  0.52~s &  73.8~s \\ \hline
Sec.~\ref{sec:Trapezoidal} & On-chip line &  0.18~s &  9.11~s \\ \hline
Sec.~\ref{Sec:Curved_Interconnect} & Curved microstrips & 0.51~s & 40.3~s \\ \hline
\end{tabular} \\
\label{tab:Timing_FEM} 
\end{table}

\begin{figure}[t]
\centering
\includegraphics{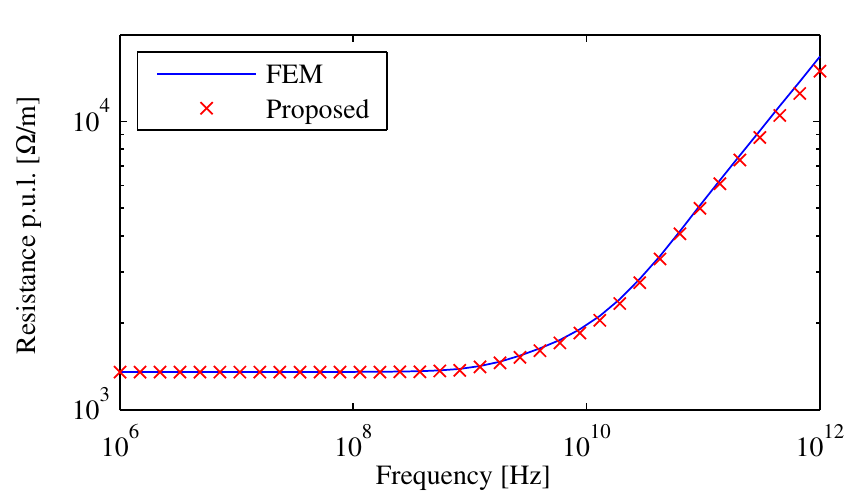}
\includegraphics{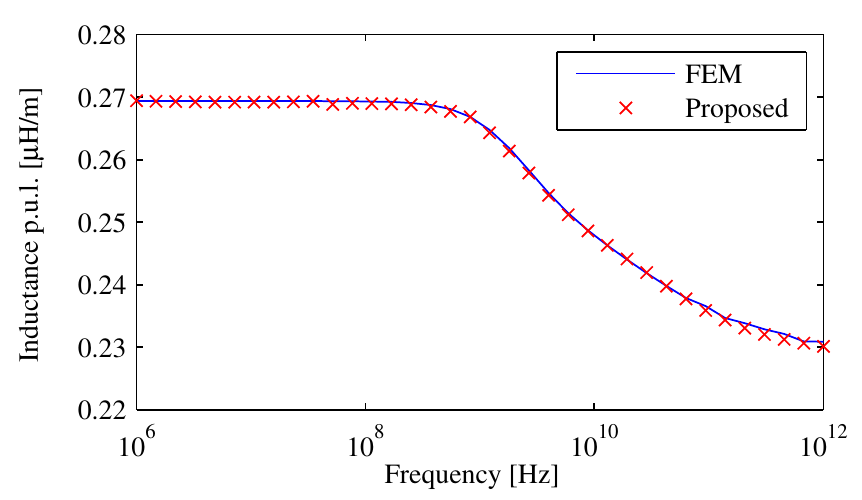}
\caption{P.u.l. resistance and inductance of the valley microstrip line considered in Sec.~\ref{sec:valley}, computed with the proposed method and the FEM~\cite{COMSOL}.}
\label{eq:valley_rl}
\end{figure}

We  consider a valley microstrip line from~\cite{Garg_2013}, having the cross-section depicted in  Fig.~\ref{fig:valley}. This line type is used in low-loss microwave integrated circuits, and has been considered in several previous publications~\cite{Rong_1993,Hasegawa_1991}.  All conductors are made of copper ($\sigma = 5.8\cdot 10^7~{\rm S/m}$). The top conductor is the signal line, while the two lower conductors form the reference line.

Figure~\ref{eq:valley_rl} shows the p.u.l. inductance and resistance of the system obtained with the FEM~\cite{Yin89,COMSOL} and with the proposed approach.
The results from the two methods agree  well both at very low (1~MHz) and very high (1~THz) frequency. This test further validates the proposed method and shows its numerical robustness. In the FEM simulation, we had to use a fine mesh with 3,102 boundary elements and 188,107 triangular elements, in order to properly resolve the pronounced skin effect at high frequency.
In the proposed technique, the boundary of the signal conductor was discretized with $N_1 = 194$ pulse basis functions, while each ground conductor was discretized with $N_2 = N_3 = 92$ pulse basis functions.
The CPU time taken by both techniques is reported in Table~\ref{tab:Timing_FEM} and shows the efficiency of the proposed method, which requires only 0.52~s to extract the p.u.l. impedance at one frequency for this non-trivial line. While in such cases one can resort to a general 2D FEM approach, this significantly increases the computational cost, as shown in Table~\ref{tab:Timing_FEM}.

\subsection{On-chip Transmission Line with Trapezoidal Conductors}
\label{sec:Trapezoidal}
\input{./TestCases/Trapezoidal/trapezoidal_geometry.tex}

\begin{figure}[t]
\centering
\includegraphics{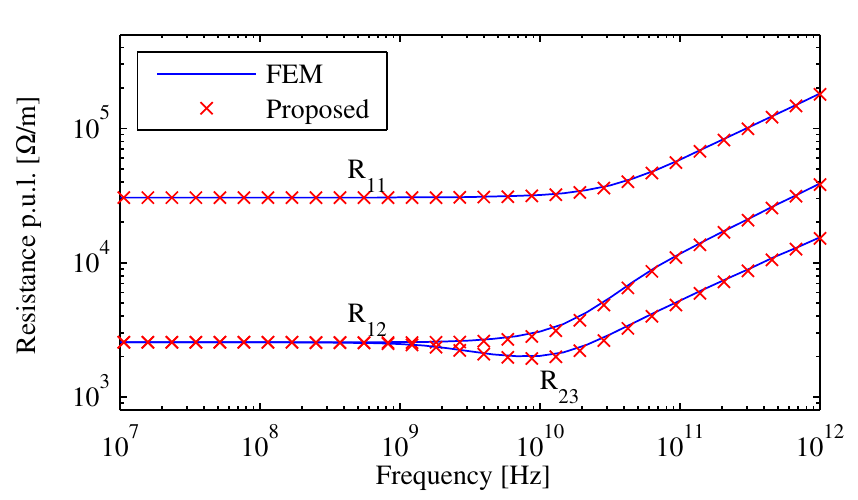}
\includegraphics{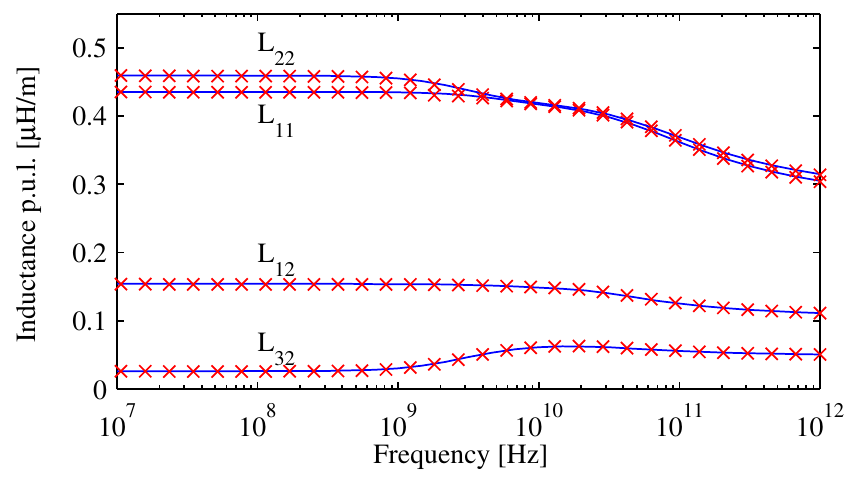}
\caption{P.u.l. resistance and inductance of the on-chip transmission line with trapezoidal conductors considered in Sec.~\ref{sec:Trapezoidal}.}
\label{fig:pul_trapezoidal}
\end{figure}

Next, we consider the four-conductor transmission line~\cite{Demeester2009} shown in Fig.~\ref{fig:trapezoidal}.
All conductors are made of Al-oxide ($\sigma = 3.57\cdot 10^7~{\rm S/m}$). This type of transmission line is typical for interconnects in integrated circuits. The trapezoidal shape of the signal lines arises is cause, for example, by  underetching or electrolytical growth. As shown in~\cite{Demeester2009}, approximating signal lines with perfect rectangles results in a non-negligible error on the p.u.l. resistance and inductance. Figure~\ref{fig:pul_trapezoidal} shows various entries of the p.u.l. resistance and inductance matrices of the transmission line, computed with the proposed technique and the FEM~\cite{COMSOL}. The proposed technique correctly captures the non-trivial impedance behaviour over frequency. In the proposed method, each trapezoidal conductor was discretized with 40 pulse basis functions, while 116 functions were used for the lower conductor. In the FEM simulation, 27,596 triangular and 718 boundary elements were used to mesh the line's cross section. Table~\ref{tab:Timing_FEM} shows that the proposed method took just 0.18~s to calculate the whole p.u.l. impedance matrix at a single frequency.

\subsection{Curved Microstrip Lines}
\label{Sec:Curved_Interconnect}
\input{./TestCases/Cylindrical_Interconnect/cylindrical_interconnect_geometry.tex}

\begin{figure}[h!]
\centering
\includegraphics{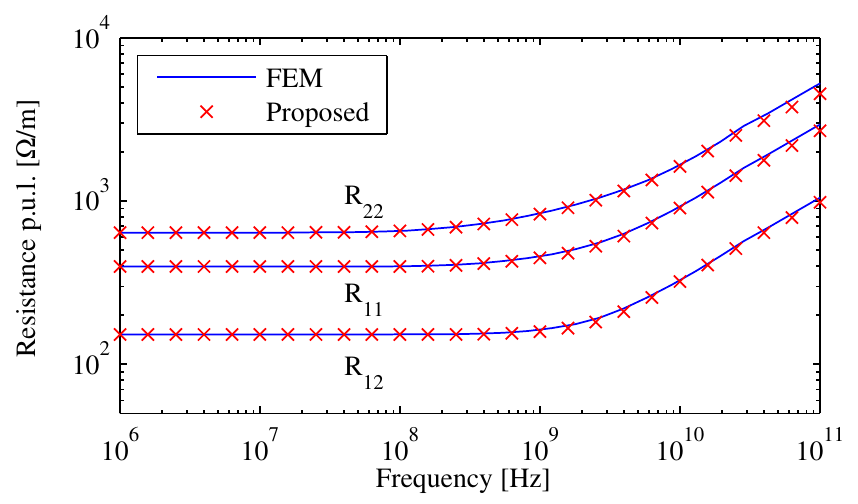}
\includegraphics{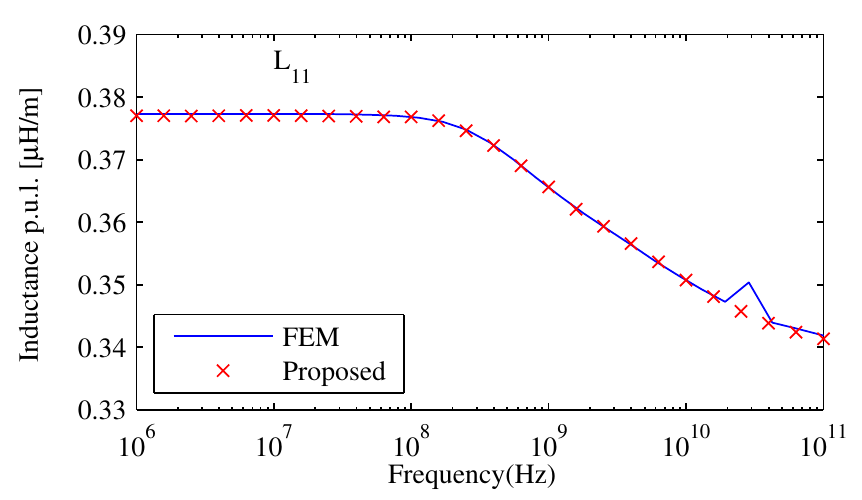}
\includegraphics{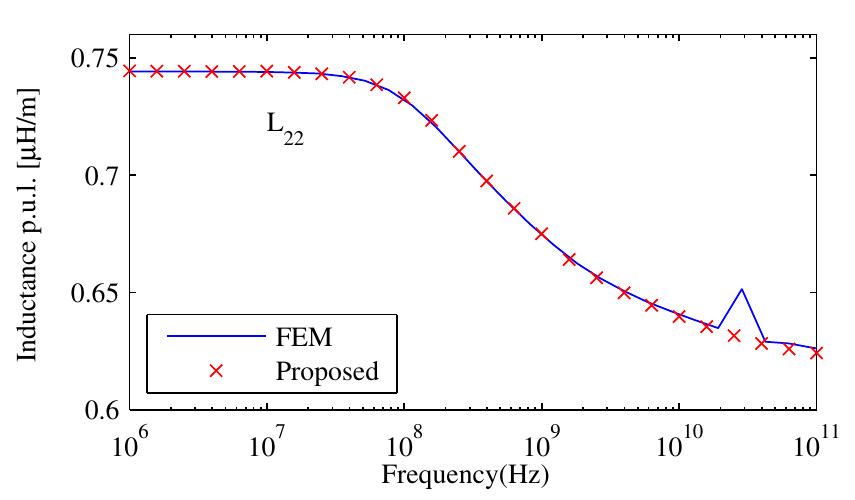}
\caption{Selected entries of the p.u.l. resistance and inductance matrices of the curved microstrip lines of Sec.~\ref{Sec:Curved_Interconnect}.}
\label{fig:RL_CurvedInterconnect}
\end{figure}

Flexible dielectrics allow for the creation of curved interconnects conformal to a cylindrical surface~\cite{Zheng_1986, Chan_1987}. We analyse the configuration shown in Fig.~\ref{fig:CurvedInterconnect_geometry}, which features four copper conductors ($\sigma = 5.8 \cdot 10^7~{\rm S/m}$). Conductors were chosen to be very thin and wide to demonstrate the robustness of the proposed approach in handling conductors with large aspect ratio. Figure~\ref{fig:RL_CurvedInterconnect} shows  selected entries of the resistance and inductance matrices, computed with the proposed method and the FEM~\cite{COMSOL}. A close match can be observed, which validates the proposed technique. Because of the low skin depth at high frequency, the FEM undergoes some numerical difficulties at high frequency in the computation of inductance, that we could not eliminate by refining the mesh. 
This issue is indicative of the intrinsic challenge in meshing the region inside the conductors at high frequency, where current is confined in a small region near the edges. Surface methods, such as~\cite{DeZ05} and the one in this paper, completely avoid this issue. In the proposed method, a total of 373 pulse basis functions were used to discretize the conductors boundary. Table~\ref{tab:Timing_FEM} shows that, with the proposed method, the p.u.l. impedance of the conformal interconnect can be obtained in 0.51~s per frequency.

%% file: TestCases/TwoRoundConductors/Twoconductor_geometry.tex
\begin{figure}[t]
\centering
\begin{tikzpicture}[scale=1]
\begin{scope}[shift={(5,0)}]
\draw [fill = black!50] (-2,-0.2) -- (-2,0.2) -- (2,0.2) -- (2,-0.2);
\draw [fill = black!50] (-2,0.8) node (v1) {} -- (-2,1.2) -- (2,1.2) -- (2,0.8);
\node at (0,-0.1) [below] {2};
\node at (2,-0.0) [right] {0.2};
\node at (0,1.2) [above] {2};
\node at (2,1) [right] {0.2};
\draw [<->] (0,0.22) -- (0,0.78);
\node at (0,0.5) [right] {0.5};
\end{scope}

\draw[fill= black!50] (-0.175*5,0.5) circle (0.5);
\draw[fill= black!50] (0.175*5,0.5) circle (0.75);
\draw [<->] (-0.175*5, 1.8) -- (0.175*5,1.80);
\draw [<->] (-0.175*5, 1.4) -- (-0.375,1.4);
\draw [<->] (0.175*5, 1.4) -- (0.125,1.4);
\node at (0.5,1.4) [above] {2};
\node at (-0.6,1.4) [above] {1};
\node at (0,1.8) [above] {3.5};
\draw [dashed] (0.175*5, 0.5) -- (0.175*5,1.8);
\draw [dashed] (0.125, 0.5) -- (0.125,1.4);
\draw [dashed] (-0.175*5, 0.5) -- (-0.175*5,1.8);
\draw [dashed] (-0.375, 0.5) -- (-0.375,1.4);






\end{tikzpicture}
\caption{Cross section of the two-conductor lines analyzed in Sec.~\ref{sec:example_twoconductor}. All dimensions are in millimeters. }
\label{fig:Twoconductor_geometry}
\end{figure}

%% file: TestCases/Valley/valley_geometry.tex
\begin{figure}
\centering
\begin{tikzpicture}[scale=0.20]
\draw [fill = black!50] (12.5,5) -- (12.5,6) -- (5,6) -- (0,1) -- (-5,6) -- (-12.5,6) -- (-12.5,5) -- (-5,5) -- (0,0) -- (5,5) -- (12.5,5);
\draw [fill = black!50] (5,-4) -- (18,-4) -- (18,-3) -- (5,-3);
\draw [fill = black!50] (-5,-4) -- (-18,-4) -- (-18,-3) node (v1) {} -- (-5,-3);
\node at (11.5,-2) {13};
\node at (-11.5,-2) {13};
\node at (18,-3.5) [right] {1};
\node at (-18,-3.5) [left] {1};
\node at (8.5,6) [above]{7.5};
\node at (-8.5,6) [above]{7.5};
\node at (12.5,5.7) [right]{1};
\node at (-12.5,5.7) [left]{1};
\node at (0,-3.5) [below]{10};
\draw [<->] (-4.75,-3.5) -- (4.75,-3.5) {};
\draw[dashed] (0,0) -- (6,0) {};
\draw[<->] (5,-2.7) -- (5,-0.3) {};
\node at (5,-1.5) [right] {3};
\node (a1) at  (18,-3) {};
\node (a2) at  (18,5) {};
\node (a3) at (12.5,5) {};
 
\draw (a1.center) -- (a2.center);
\draw (a3.center) -- (a2.center);

\node (a4) at (-18,-3) {};
\node (a6) at (-18, 5) {};
\node (a5) at (-12.5,5) {};

\draw  (a4.center) edge (a6.center);
\draw (a6.center) edge (a5.center);
\node (a7) at (-5,-3) {};
\node (a8) at (5,-3) {};
\draw (a7.center) edge (a8.center);
\end{tikzpicture}
\caption{Valley microstrip line considered in Sec.~\ref{sec:valley}. All dimensions are in micrometers. }
\label{fig:valley}
\end{figure}

%% file: TestCases/Trapezoidal/trapezoidal_geometry.tex
\begin{figure}[t]
\centering
\begin{tikzpicture}[scale=0.6, every node/.style={scale=0.6}]
\draw [fill = black!50] (-1.25,1.5) -- (-1.75,2.5) -- (-2.25,2.5) -- (-2.75,1.5);
\draw [fill = black!50] (0.75,1.5) -- (0.25,2.5) -- (-0.25,2.5) -- (-0.75,1.5);
\draw [fill = black!50] (2.75,1.5) -- (2.25,2.5) -- (1.75,2.5) -- (1.25,1.5);
\draw  (-5.5,0.5)  rectangle (5.5,1.5);
\draw [fill = black!50] (-5.5,-0.5) -- (-5.5,0.5) node (v1) {} -- (5.5,0.5) -- (5.5,-0.5);

\node at (0,-0.5) [below]{\Large 13};
\node at (5.5,0) [right]{\Large 1};
\node at (0,1.5) [below]{\Large 1.5};
\node at (0,2.5) [above]{\Large 0.75};

\draw [dashed] (-2,2.5) -- (4.5,2.5);
\draw [dashed] (3.5,1.5) -- (4.5,1.5);
\draw [<->] (4,2.4) -- (4,1.6);
\draw [<->] (0.8,1.4) -- (1.20,1.4);
\draw [<->] (4,1.4) -- (4,0.6);
\node at (4,1) [right]{\Large 1};
\node at (1,1.4) [below]{\Large 0.5};

\node at (4,2) [right] {\Large 1};
\node [color = black] at (0,0) {\Large \#0};
\node [color = black] at (0,2) {\Large \#1};
\node [color = black] at (2,2) {\Large \#2};
\node [color = black] at (-2,2) {\Large \#3};

\node at (-4.5,1) {\Large $\varepsilon_{0}, \mu_0$};
\end{tikzpicture}
\caption{On-chip interconnect of Sec.~\ref{sec:Trapezoidal}. All dimensions are in micrometers. }
\label{fig:trapezoidal}
\end{figure}

%% file: TestCases/Cylindrical_Interconnect/cylindrical_interconnect_geometry.tex
\begin{figure}[t]
\centering
\begin{tikzpicture}[scale=0.7, every node/.style={scale=0.7}]

\draw[fill=black!50,fill opacity=0.7] ([shift=(110:3)]0,0) arc (110:70:3) --++ (70:0.3) arc (70:110:3.3) -- cycle; 
\draw[fill=black!50,fill opacity=0.7] ([shift=(100:4)]0,0) arc (100:80:4) --++ (80:0.3) arc (80:100:4.3) -- cycle; 

\draw[fill=black!50,fill opacity=0.7] ([shift=(65:4)]0,0) arc (65:55:4) --++ (55:0.3) arc (55:65:4.3) -- cycle; 
\draw[fill=black!50,fill opacity=0.7] ([shift=(125:4)]0,0) arc (125:115:4) --++ (115:0.3) arc (115:125:4.3) -- cycle; 
 \draw[fill = black!80]  (0,0) ellipse (0.05 and 0.05);
 \node (p1) at (90:5) {} ;
 \node (p2) at (80:5) {};
 
 \node (p4) at (60:5) {};
 \node (p5) at (55:5) {};
 \node (p6) at (70:5) {}; 
 
 \node (origin) at (0,0) {};
 \draw [dashed, color = black!50] (p1) -- (origin.center);
 \draw [dashed, color = black!50] (origin.center) -- (p2);
 
 \draw [dashed, color = black!50] (origin.center) -- (p4);
 \draw [dashed, color = black!50] (origin.center) -- (p5);
 \draw [dashed, color = black!50] (origin.center) -- (p6);

 \draw[->]  ([shift=(90:1.4)]0,0) arc(90:80:1.4);
 \node at (85:1.2) {\small $\alpha_1$};
 \draw[->]  ([shift=(90:1.8)]0,0) arc(90:70:1.8);
 \node at (80:1.6) {\small $\alpha_2$};
 \draw[->]  ([shift=(90:2.1)]0,0) arc(90:60:2.1);
 \node at (75:1.95) {\small $\alpha_3$};
 \draw[->]  ([shift=(90:2.4)]0,0) arc(90:55:2.4);
 \node at (75:2.28) {\small $\alpha_4$};
\draw (4,0) arc (0:180:4) {};  
\draw (3,0) arc (0:180:3) {};
\node (p7) at (20:3) {};
\node (p8) at (35:4) {};
\draw[->] (origin.center) -- (p7.center) {};
\draw[->] (origin.center) -- (p8.center) {};
\node at (20:1.5) [below] {$R_1$};
\node at (37:2) [above] {$R_2$};
\node (p9) at (112:3) {};
\node (p10) at (112:3.3){};
\node (p11) at (127:4) {};
\node (p12) at (127:4.3) {};
\node (p13) at (102:4){};
\node (p14) at (102:4.3){};
\draw[<->] (p9.center) -- (p10.center);
\draw[<->] (p11.center) -- (p12.center);
\draw[<->] (p13.center) -- (p14.center);
\node at (112:3.15) [left]{$h$};
\node at (102:4.15) [left]{$h$};
\node at (127:4.15) [left]{$h$};
\node [color = black] at (90:3.15) {\#0};
\node [color = black] at (90:4.15) {\#1};
\node [color = black] at (60:4.15) {\#2};
\node [color = black] at (120:4.15) {\#3};
\end{tikzpicture}
\caption{Cross-section of the curved microstrips of Sec.~\ref{Sec:Curved_Interconnect}. Geometrical dimensions are: $\alpha_1 = 10^\circ$, $\alpha = 20^\circ$, $\alpha = 30^\circ$, $\alpha = 35^\circ$, $R_1 = 80~\mu{\rm m}$, $R_2 = 100~\mu{\rm m}$, $h = 2~\mu{\rm m}$.}
\label{fig:CurvedInterconnect_geometry}
\end{figure}

%% file: Conclusion.tex
\section{Conclusion}

We presented an efficient approach to
model skin effect in conductors of arbitrary cross section. The proposed approach is based on a surface admittance operator. We show how the operator can be efficiently computed from a contour integral. The proposed approach can handle conductors of arbitrary shape, unlike the popular eigenfunctions method which is viable only for canonical shapes, where eigenfunctions are available analytically. The novel method has been applied to the computation of the resistance and inductance of transmission lines with rectangular, circular, trapezoidal, V-shaped and curved conductors. Numerical results demonstrate the efficiency, robustness and accuracy of the proposed technique.